\documentclass[11pt]{article}
\usepackage{latexsym}
\usepackage{amssymb}
\usepackage{epsfig}
\usepackage{rotating}
\usepackage{amsfonts}
\usepackage{amsmath}
\newcommand{\rmd}{\mathrm{d}}
\begin{document}

\begin{titlepage}
\begin{flushright}
{\  }CP3-18-40\\
\end{flushright}

\vspace{10pt}

\begin{center}

{\Large\bf Mass Generation in Abelian U(1) Gauge Theories:}\\

\vspace{5pt}

{\Large\bf A Rich Network of Dualities}\\

\vspace{40pt}

Bruno Bertrand$^{a}$\footnote{ORCID: {\tt https://orcid.org/0000-0003-4538-1452}} and 
Jan Govaerts$^{b,c,}$\footnote{Fellow of the Stellenbosch Institute for Advanced Study (STIAS),
Stellenbosh, Republic of South Africa}$^{,\!\!\!}$
\footnote{ORCID: {\tt http://orcid.org/0000-0002-8430-5180}}

\vspace{20pt}

$^{a}${\sl Royal Observatory of Belgium (ROB),
3, Avenue Circulaire, 1180 Brussels, Belgium}\\
E-mail: {\em Bruno.Bertrand@oma.be}

\vspace{10pt}

$^{b}${\sl Centre for Cosmology, Particle Physics and Phenomenology (CP3),\\
Institut de Recherche en Math\'ematique et Physique (IRMP),\\
Universit\'e catholique de Louvain (UCL),\\
2, Chemin du Cyclotron, bte L7.01.01, B-1348 Louvain-la-Neuve, Belgium}\\
E-mail: {\em Jan.Govaerts@uclouvain.be}

\vspace{10pt}

$^{c}${\sl International Chair in Mathematical Physics and Applications (ICMPA--UNESCO Chair),\\
University of Abomey--Calavi, 072 B.P. 50, Cotonou, Republic of Benin}

\vspace{20pt}


\vspace{20pt}


\begin{abstract}
\noindent
Following a novel approach, all known basic mass generation mechanisms consistent with an exact abelian U(1)
gauge symmetry are shown to be related through an intricate network of dualities whatever the spacetime dimension.
This equivalence which applies in the absence of any supersymmetry, is however restricted by the presence
of topological terms generating possible topological effects. In particular in $3+1$ dimensions
the duality relations between the Maxwell-Higgs model, the Stueckelberg and the topological mass generation
mechanisms are then established following a careful treatment of the gauge symmetry content. This result offers
a new framework for an effective description of superconductivity or topological defects built from fields beyond the SM. 

\end{abstract}

\end{center}

\end{titlepage}

\setcounter{footnote}{0}

\section*{Introduction and Overview}

\begin{figure}[h]
\begin{center}
\includegraphics[width=13cm]{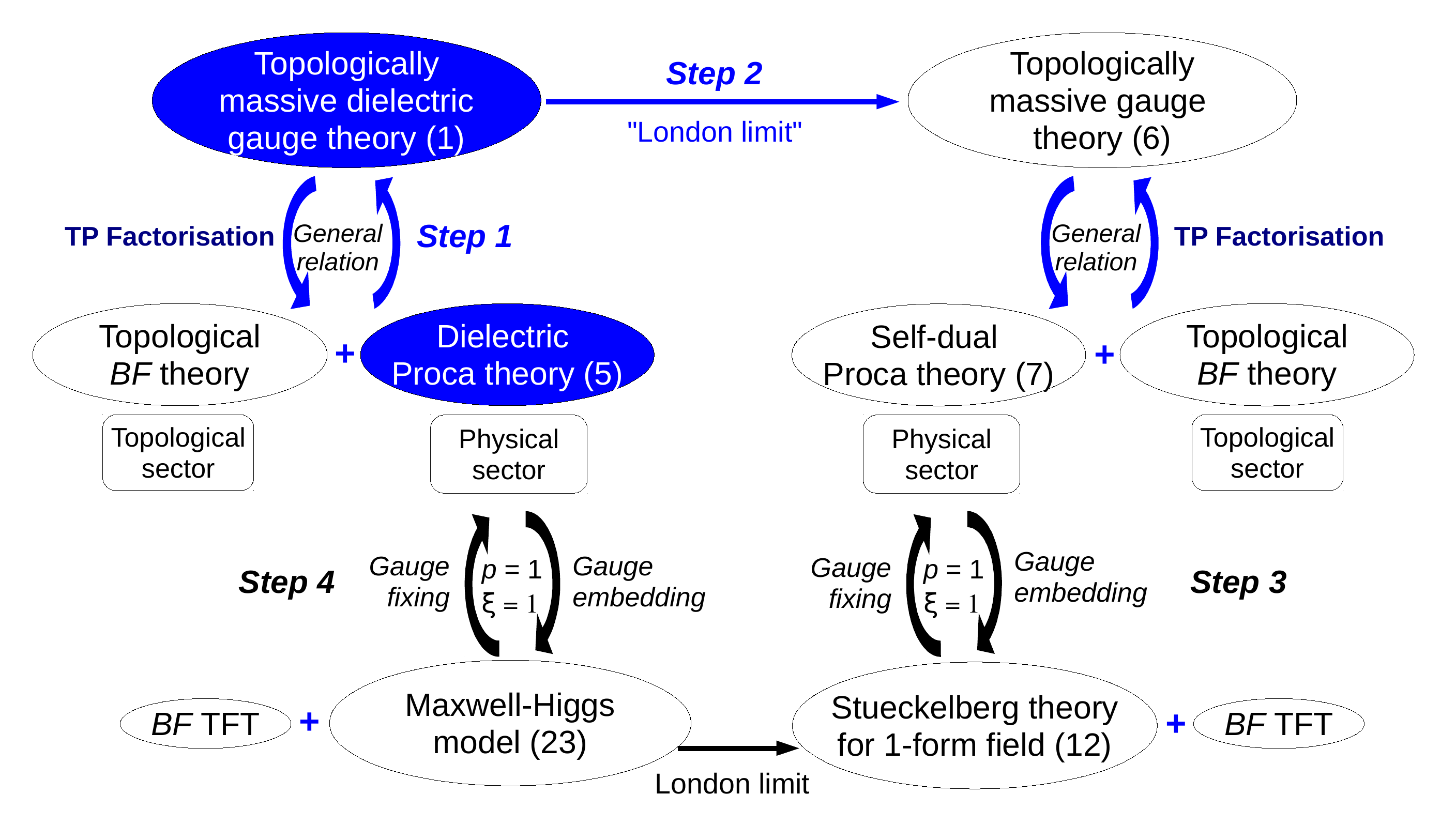}
\caption[]{\textit{Network of (local) duality relations between mass generation mechanisms in abelian U(1)
gauge field theories. In blue: our contributions.}}
\label{fig:Duality_Mass-generation}
\end{center}
\end{figure}

Among a number of fundamental concepts, the understanding of the origin of mass and the related formation of topological defects remains among those open problems of great relevance in particle physics, cosmology
or condensed matter. Within this context the present contribution sheds new light on mass generation mechanisms
through the notion of duality which is known to play a crucial role in high energy physics. Its main purpose is
to establish the network of dualities displayed in Fig.\ref{fig:Duality_Mass-generation} where all the celebrated mass generation mechanisms consistent with an exact abelian U(1) gauge symmetry are recovered but now find their rightful place. Each equivalence is defined modulo the presence of topological terms generating possible topological effects.

The construction of this network of duality relations is framed in four steps which will be described in detail.
This includes the different duality techniques used along with the terminology characterizing the different
intermediate Lagrangian densities. Our analysis is restricted to the abelian cases, used as laboratories
for mass gap generation where important effects of topological origin are already manifest. However
in the present discussion it will be assumed that the fields do not vanish anywhere on the topologically
trivial spacetime manifold, thereby avoiding considerations related to the possibility of topological defects.
The results presented herein were originally first established in \cite{BB-PhD}.

\section*{Step One - Generalisation of TP factorisation to TMDGT}

Topologically massive gauge theories (TMGT) are dynamical field theories in which a topological term preserving the abelian U(1) gauge invariance generates a mass gap, see \cite{Cremmer:1973mg,Hagen:1978jk,Allen:1990gb} in 3+1 dimensions. A natural extension of the general action for TMGT in a $d\!+\!1$ dimensional Minkowski
spacetime as introduced in \cite{Bertrand:2007ay}, is obtained by transmuting the gauge field normalisation factors\footnote{Or gauge coupling constants when coupled to matter fields.} $e$ and $g$ into functions
of dynamical real scalar fields
$\varpi(x)$ and $\varrho(x)$. Hence the resulting action, $S^{d+1}_{\mathrm{TM}\varrho\varpi}[A,B,\varrho,\varpi]$, reads
\begin{eqnarray} \label{def:TMGT+Dielectric_fields}
S^{d+1}_{\mathrm{TM}\varrho\varpi} & = & \frac{1}{2} \int_{\mathcal{M}} \frac{\sigma^{p}}{e^2(\varpi)} F\wedge\ast F + \frac{\sigma^{d-p}}{g^2(\varrho)} H\wedge\ast H \nonumber\\ & + & S_{BF}[A,B] + \mathcal{S}_{\varpi}\left[\varpi\right] + \mathcal{S}_{\varrho}\left[\varrho\right]\, ,
\end{eqnarray}
where the notations of \cite{Bertrand:2007ay} are being used throughout. The U(1) gauge invariant 
fields $\varpi(x)$ and $\varrho(x)$ couple through the dielectric functions $e\left(\varpi(x)\right)$ and $g\left(\varrho(x)\right)$ to the kinetic terms of the $p$-form $A(x)$ and the ($d\!-\!p$)-form $B(x)$, respectively. Hence such types of actions will be called topologically massive dielectric gauge theories (TMDGT). In the present discussion
it is assumed that $\varpi(x)$ and $\varrho(x)$ do not vanish anywhere on the spacetime manifold $\mathcal{M}$.
Within the topological coupling $S_{BF}[A,B]$ appearing in (\ref{def:TMGT+Dielectric_fields}),
\begin{displaymath}
S_{BF}[A,B] = \kappa\, \int_{\mathcal{M}} (1-\xi)\, F\wedge B - \sigma^p\, \xi\, A\wedge H\, ,
\end{displaymath}
the arbitrary real variable $\xi \in [0,1]$ is physically irrelevant (at the classical level) for an appropriate choice
of boundary conditions, and parametrises a partial integration by parts.

The Topological-Physical factorisation technique consists in constructing a dual formulation of TMGT which is factorised into a dynamical sector of massive physical (i.e. gauge invariant) variables and a gauge dependent sector defining a topological field theory (TFT). Our factorisation introduced in \cite{BB-PhD,Bertrand:2007ay} is a particular type of dual projection technique (see for example \cite{Abreu:2007pp} for a review) which turns out to be the covariant extension
of an equivalent canonical transformation within the Hamiltonian formulation. Interestingly the action (\ref{def:TMGT+Dielectric_fields}) is the most general construction of topologically massive gauge fields coupled to real scalar fields of which the decoupled part is minimal,
\begin{eqnarray}
\mathcal{S}_{\varrho}\left[\varrho\right] & = & \int_{\mathcal{M}} \frac{1}{2}\, \rmd\varrho\wedge\ast\rmd\varrho - \ast V\left( \varrho^2 \right) \, , \label{def:varrho_action}\\
\mathcal{S}_{\varpi}\left[\varpi\right] & = & \int_{\mathcal{M}} \frac{1}{2}\, \rmd\varpi\wedge\ast\rmd\varpi - \ast \tilde{V}\left( \varpi^2 \right) \, , \label{def:varpi_action}
\end{eqnarray}
and which at the same time preserves our factorisation into decoupled physical and topological sectors. 
There is no need at this stage to specify the potentials $\tilde{V}\left(\varpi^2 \right)$ and $V\left(\varrho^2 \right)$.

In spite of the presence of dielectric fields in (\ref{def:TMGT+Dielectric_fields}), the dual factorised formulation is obtained in the same way as within the non coupled case, see \cite{BB-PhD,Bertrand:2007ay},
and thus proceeds first through the extension of the field content. Indeed the covariant first order formulation of (\ref{def:TMGT+Dielectric_fields}) reads
\begin{eqnarray}
 S^{\mathrm{master}}_{\mathrm{TM}\varrho\varpi} & = & e^2(\varpi) \frac{E^2}{2} + g^2(\varrho) \frac{G^2}{2} + \mathcal{S}_{\varpi}\left[\varpi\right] + \mathcal{S}_{\varrho}\left[\varrho\right] \nonumber\\
 & + & \int_{\mathcal{M}} F \wedge E + H \wedge G + S_{BF}[A,B] \nonumber\, ,
\end{eqnarray}
after the introduction of the auxiliary $(d\!-\!p)$- and $p$-form fields $E(x)$ and $G(x)$, see \cite{Bertrand:2007ay} for the meaning
of notations. Then through a reparametrisation similar to that of the non coupled case,
\begin{equation}\label{def:TP-factorisation}
A = \mathcal{A} + \frac{1}{\kappa} \sigma^{p(d-p)}\, G \quad , \quad B = \mathcal{B} - \frac{1}{\kappa} E \, ,
\end{equation}
the first order action $S^{\mathrm{master}}_{\mathrm{TM}\varrho\varpi}$ factorises as follows, 
\begin{displaymath}
S^{\mathrm{fac}}_{\varrho\varpi} = S_{\mathrm{dyn}}[E,G,\varrho,\varpi] + S_{BF}[\mathcal{A},\mathcal{B}] + \int_{\mathcal{M}} \mathrm{ST} \, ,
\end{displaymath}
into a $BF$ TFT, $S_{BF}$, and a dynamical sector $S_{\mathrm{dyn}}$ of the form
\begin{eqnarray}\label{def:TMGT+Diel_dyn}
S_{\mathrm{dyn}} & = & e^2(\varpi) \frac{E^2}{2} + g^2(\varrho) \frac{G^2}{2} + \mathcal{S}_{\varpi}\left[\varpi\right] + \mathcal{S}_{\varrho}\left[\varrho\right] \nonumber\\
& + & \int_{\mathcal{M}} \sigma^{d-p}\, \frac{\xi}{\kappa} E \wedge \rmd G - \frac{1-\xi}{\kappa}\, \rmd E \wedge G \, , 
\end{eqnarray}
up to a surface term, $ST$, irrelevant for an appropriate choice of boundary conditions on $\mathcal{M}$.
This new general action for the physical sector, $S_{\mathrm{dyn}}$, is referred to as ``dielectric Proca theory'' in Fig.\ref{fig:Duality_Mass-generation}. 

\section*{Step Two - The ``London limit'' for TMDGT}

In the Maxwell-Higgs model, the London limit is the limit in which the mass of the Higgs field becomes infinite while the mass of the gauge field remains finite. As a matter of fact, within this limit the dynamics of the Higgs field is frozen to its vacuum expectation value. The action resulting from the decoupling of the massive Higgs field is nothing other than the Stueckelberg action for a 1-form gauge field (see \cite{Ruegg:2003ps} and references therein).

Likewise an equivalent limit may be readily identified for the TMDGT defined in (\ref{def:TMGT+Dielectric_fields}), within the context of the duality relations displayed in Fig.\ref{fig:Duality_Mass-generation}. However, as there is no need to specify the shape of the self-interacting scalar potentials in (\ref{def:varrho_action}) and (\ref{def:varpi_action}), by ``London limit'' we shall refer to any asymptotic limit or relation between the coupling constants which leads to the ``freezing'' of the scalar fields to their vacuum expectation values. In this sense, the general action for topologically massive gauge theories in \cite{BB-PhD,Bertrand:2007ay}, 
\begin{equation} \label{def:TMGT_Action}
S^{d+1}_{\mathrm{TM}} = \int_{\mathcal{M}} \frac{\sigma^{p}}{2 \, e^2}\, F\wedge\ast F + \frac{\sigma^{d-p}}{2 \, g^2} \, H\wedge\ast H + S_{BF}\, ,
\end{equation}
is recovered through the London limit,
\begin{displaymath}
e\left(\varpi\right)\, \to\, e\left(\left\langle \varpi \right\rangle\right) \equiv e\, , \qquad
g\left(\varrho\right)\, \to\, g\left(\left\langle \varrho \right\rangle\right) \equiv g\, ,
\end{displaymath}
of the action (\ref{def:TMGT+Dielectric_fields}) for TMDGT.

\section*{Step Three - Equivalence with Stueckelberg theories revisited}
\label{sec:dual_Stueckelberg}

In fact, several dualisation techniques have been used until now in order to establish the dual equivalence between topologically massive gauge theories and Stueckelberg theories. As far as gauge embedding procedures are concerned, this duality relation is considered as an intermediate step in order to establish the duality relation between theories of the Proca-type and TMGT. This type of methods, developed whether within the Hamiltonian \cite{Kim:2003ux} or the Lagrangian \cite{Menezes:2003vz,Hansson:2004wca} formulation, are characterised by an intricate maze of successive gauge fixing and unfixing procedures. There also exist other techniques of dualisation through master Lagrangian approaches, see \cite{Lee:1993ty,Smailagic:1999qw,Cantcheff:2001ws}. Even in this latter case however no special care has been taken with regards to the gauge symmetry content and its possible non trivial topological properties.

In contradistinction to these procedures, our method consists of two steps, as illustrated in the right-hand part of Fig.\ref{fig:Duality_Mass-generation}. First we have already established in \cite{Bertrand:2007ay}, through our TP factorisation, the duality relation between TMGT (\ref{def:TMGT_Action}) and a generalised first order formulation of Proca theories, 
\begin{eqnarray}\label{def:Generalised_Proca}
S_{\mathrm{dyn}} & = & e^2\, \frac{E^2}{2} + g^2\, \frac{G^2}{2} \\
 & + & \int_{\mathcal{M}} \sigma^{d-p}\, \frac{\xi}{\kappa} E \wedge \rmd G - \frac{1-\xi}{\kappa}\, \rmd E \wedge G \, , \nonumber
\end{eqnarray}
modulo a topological $BF$ term in which all the gauge content resides. This procedure is free of any gauge fixing
procedure whatsoever and is fully consistent for what concerns the counting of the numbers of degrees of freedom, inclusive of topological ones and pure gauge ones. Note that the action (\ref{def:Generalised_Proca}) is the non gauge invariant ``self-dual'' action of \cite{Menezes:2003vz,Cantcheff:2001ws} in any dimension, recovered
upon setting $\xi\!=\!0$.

Second it is only at this stage that gauge embedding procedures apply in order to make manifest the duality between the Proca theories in the physical sector and the Stueckelberg theories.
Let us briefly recall the basic concepts of this type of generic procedure consisting in the extension of the gauge content. Knowing that $\rmd^{\dag}{G}\!=\!0$ and $\rmd^{\dag}{E}\!=\!0$ on shell, the two gauge invariant co-closed $p$ and $(d\!-\!p)$-form fields $G(x)$ and $E(x)$ may be written as
\begin{eqnarray}
G & = & \sigma^{p(d-p)}\, \kappa \left( \tilde{A} - \theta \right) \, , \label{def:Gauge_Embedding-G}\\
E & = & - \kappa \left( \tilde{B} - \chi \right)\, , \label{def:Gauge_Embedding-E}
\end{eqnarray}
where $\theta(x)$ is a closed $p$-form while $\chi(x)$ is a closed $(d\!-\!p)$-form.

In contradistinction to what was so far advocated in the literature, we prefer to extend the field (and gauge) content within the gauge embedding procedure rather than to involve again the original gauge fields $A(x)$ and $B(x)$.  Indeed let us imagine that we set $\tilde{A}(x) = A(x)$ and $\tilde{B}(x) = B(x)$ quite similarly to what was done in \cite{Cremmer:1973mg,Smailagic:1999qw,Harikumar:1999tm}. Then using the reparametrisations (\ref{def:Gauge_Embedding-G}) and (\ref{def:Gauge_Embedding-E}) in combination with the transformations (\ref{def:TP-factorisation}) the following constraints are obtained
\begin{eqnarray}
  \mathcal{A} (x) = \theta (x) & \textrm{if} & \tilde{A}(x) = A(x) \, ,\nonumber\\ 
  \mathcal{B} (x) = \chi (x) & \textrm{if} & \tilde{B}(x) = B(x) \, . \nonumber
\end{eqnarray}
Actually this choice is inconsistent with the original counting of the total number of degrees of freedom since it sets the transverse part of $\mathcal{A}(x)$ and $\mathcal{B}(x)$ to zero.  Hence a first-class constraint would already be satisfied within the transformation itself rather than on shell.

Under the restrictive assumptions of a topologically trivial spacetime manifold,
$\theta(x)$ and $\chi(x)$ are also exact\footnote{These assumptions do not allow $\theta(x)$ and $\chi(x)$ to have a global component, see \cite{Nielsen:1973cs,Lee:1993ty} in 3+1 dimensions, and thus exclude the existence of topological defect solutions.}.
In this case, the gauge embedding procedure is then well-defined. The transformations of the fields $\theta(x)$ and $\chi(x)$ under the new gauge symmetries, 
\begin{displaymath}
\theta' = \theta + \tilde{\alpha} \, , \qquad  \chi' = \chi + \tilde{\beta} \, ,
\end{displaymath}
where $\tilde{\alpha}(x)$ and $\tilde{\beta}(x)$ are two exact $p$- and $(d\!-\!p)$-forms, compensate for those of the two independent classes of abelian gauge transformations acting separately in either the $\tilde{A}$- or $\tilde{B}$-sector,
\begin{eqnarray}
\tilde{A}' & = & \tilde{A} + \tilde{\alpha} \, , \label{def:Gauge_embedding-A'}\\
\tilde{B}' & = & \tilde{B} + \tilde{\beta}\, , \label{def:Gauge_embedding-B'}
\end{eqnarray}
in order to preserve the gauge invariance of $G(x)$ and $E(x)$. In this sense the physical variable $G(x)$ may be considered as the gauge invariant transverse part of the gauge field variable $\tilde{A}(x)$ while $\theta(x)$ is associated to its longitudinal part. A likewise identification applies for the $(d\!-\!p)$-form field $B(x)$.

The dual gauge embedded Stueckelberg theory constructed from the physical sector of the factorised TMGT depends dramatically on the value of $\xi$. 
Indeed, let first set $\xi\!=\!1$ in the factorised Lagrangian density (\ref{def:Generalised_Proca}). Integrating out the then Gaussian auxiliary $(d\!-\!p)$-form field $E(x)$ and extending the gauge content of the theory at the level of the $p$-form field $G(x)$ through the transformation (\ref{def:Gauge_Embedding-G}), one derives the Stueckelberg action of a $p$-form field $\tilde{A}(x)$,
\begin{equation}\label{def:Stueck_p-form(2)}
S_{\mathrm{dyn}} = \frac{\sigma^{p}}{2 \, e^2}\, \int_{\mathcal{M}} \tilde{F}\wedge\ast \tilde{F} + \frac{\mu^2}{2\, e^2} \left( \tilde{A} - \theta \right)^2 \, ,
\end{equation}
where $\tilde{F}\!=\!\rmd \tilde{A}$ and $\mu\!=\!\kappa\, e\, g$. Alternatively the Stueckelberg action of a $(d\!-\!p)$-form field is obtained by fixing $\xi\!=\!0$, eliminating the Gaussian field $G(x)$ and applying (\ref{def:Gauge_Embedding-E}),
\begin{equation}\label{def:Stueck_(d-p)-form(2)}
S_{\mathrm{dyn}} =  \frac{\sigma^{d-p}}{2 \, g^2}\, \int_{\mathcal{M}} \tilde{H}\wedge\ast \tilde{H} + \frac{\mu^2}{2\, g^2} \left( \tilde{B} - \chi \right)^2 \, ,
\end{equation}
where $\tilde{H}\!=\!\rmd \tilde{B}$. These two actions are invariant under one abelian gauge symmetry only, of which the associated transformation reads as in (\ref{def:Gauge_embedding-A'}) or (\ref{def:Gauge_embedding-B'}) whether one considers the transformation acting on $\tilde{A}(x)$ in (\ref{def:Stueck_p-form(2)})
or on $\tilde{B}(x)$ in (\ref{def:Stueck_(d-p)-form(2)}), respectively.

\begin{figure}
\begin{center}
\includegraphics[width=13cm]{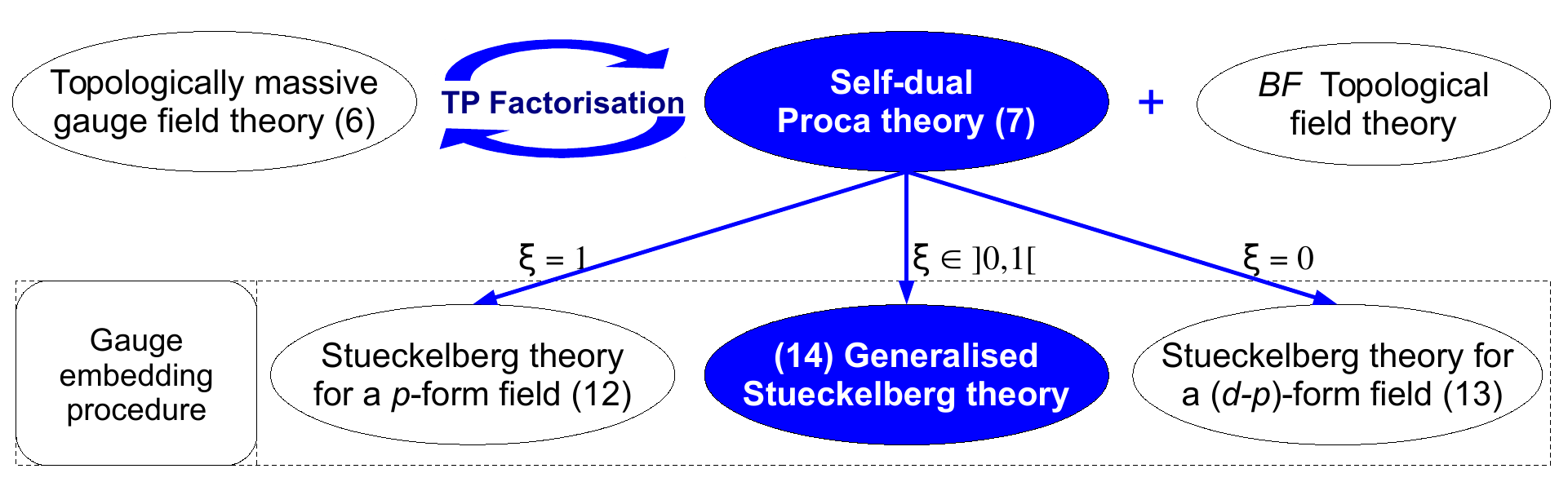}
\caption{\textit{Duality relation between TMGT (\ref{def:TMGT+Dielectric_fields}) and a generalised formulation of the Stueckelberg mechanism in abelian gauge field theories. In blue: our contributions.}}
\label{fig:Duality_Stueckelberg}
\end{center}
\end{figure}

Finally a generalised formulation of the Stueckelberg theory is obtained starting from (\ref{def:Generalised_Proca}) and applying the gauge embedding procedure to the gauge invariant fields $G(x)$ and $E(x)$, see (\ref{def:Gauge_Embedding-E}) and (\ref{def:Gauge_Embedding-G}),
\begin{eqnarray}\label{def:Generalised_Stueckelberg}
S_{\rm{dyn}} & = & \frac{\kappa^2 e^2}{2} \left( \tilde{B} - \chi \right)^2 + \frac{\kappa^2 g^2}{2} \left( \tilde{A} - \theta \right)^2 \\
& + & \kappa\, \int_{\mathcal{M}} \sigma^{d-p}\, (1-\xi)\, \tilde{A} \wedge \rmd \tilde{B} - \xi\, \rmd \tilde{A} \wedge \tilde{B} \, . \nonumber
\end{eqnarray}
Hence Fig.\ref{fig:Duality_Stueckelberg} illustrates how $\xi$ determines to which (gauge embedded) Stueckelberg theory the TMGT are dual\footnote{The analysis as developed so far in the literature considered the duality relations between TMGT and the different types of Stueckelberg theories as distinct cases \cite{Smailagic:1999qw,Harikumar:1999tm}.}. This generalised formulation of the Stueckelberg theory possesses two independent classes of abelian gauge transformations acting separately in either the $\tilde{A}$- or the $\tilde{B}$-sector, see (\ref{def:Gauge_embedding-A'}) and (\ref{def:Gauge_embedding-B'}).

\section*{Step Four - Recovering the Maxwell-Higgs model}

Finally, the range of dual formulations for Maxwell-Higgs (MH) models in $d\! +\!1$ dimensions
pertains to a very specific form of topological mass generation for a 1-form gauge field $A(x)$ with a dielectric coupling between a real scalar and the complementary ($d\! -\!1$)-form field $B(x)$. Hence starting from the action for TMGT in (\ref{def:TMGT+Dielectric_fields}) subjected to the following restrictions:
\begin{itemize}
 \item One of the gauge fields is a $1$-form field. Let us choose $A$ to be this field,
 \item The London limit applies to the field $\varpi(x)$ and the dielectric function $e\left(\varpi\right)$ thus reduces to a scaling constant $e$, 
 \item The dielectric function $g\left(\varrho(x)\right)$ has the simple form : $g(\varrho) = \eta\, \varrho$,
\end{itemize}
the following dual formulation of the MH model is obtained:
\begin{eqnarray}\label{def:4D-TMGT+Higgs}
\mathcal{L}^4_{\mathrm{TM}\varrho} & = & \frac{-1}{4\, e^2} F_{\mu\nu}\, F^{\mu\nu} + \frac{1}{12\, \eta^2} \frac{1}{\varrho^2} H_{\mu\nu\rho} \, H^{\mu\nu\rho} + \mathcal{L}_{\varrho} \nonumber\\
& + & \kappa \epsilon^{\mu\nu\rho\sigma} \left( \frac{\xi}{6} A_{\mu} H_{\nu\rho\sigma} + \frac{1-\xi}{4} F_{\mu\nu} B_{\rho\sigma} \right)\, . 
\end{eqnarray}
where $\eta$ is a normalisation parameter of physical dimension $E\, L$. For the sake of simplicity this fourth step of the structure of Fig.\ref{fig:Duality_Mass-generation} is described in $3\! +\! 1$ dimensions.
As usual, $\mathcal{L}_{\varrho}$ is the Lagrangian density for a massive real scalar field:
\begin{equation}\label{def:Free_Higgs(2)}
   \mathcal{L}_{\varrho}\left(\varrho, \partial_\mu \varrho \right) = \frac{1}{2} \partial_\mu \varrho\, \partial^\mu \varrho - V\left(\varrho^2\right) \, ,
\end{equation}
where the self-interaction potential $V\left(\varrho^2\right)$ remains unspecified at this stage.

We have already proved without having recourse to any gauge fixing choice that $\xi$ parame\-tri\-ses a classification of generalised Stueckelberg theories. Likewise this parameter offers an elegant interpretation of (\ref{def:4D-TMGT+Higgs}) in terms of the field $A_{\mu}(x)$ or the field $B_{\mu\nu}(x)$ whether $\xi\!=\!1$ or $\xi\!=\!0$, respectively. Let us consider the case $\xi\!=\!1$ and isolate the contributions involving the gauge field $A_{\mu}(x)$,
\begin{equation}\label{def:4D-TMGT+Higgs_xi=1}
\mathcal{L}^{\xi=1}_{\mathrm{TM}\varrho} = \frac{1}{12\, \eta^2} \frac{1}{\varrho^2} H_{	\mu\nu\rho} \, H^{\mu\nu\rho} + \mathcal{L}_{\varrho} + \mathcal{L}^A_{\mathrm{TM}\varrho}\, .
\end{equation}
This Lagrangian density was obtained previously by K. Lee \cite{Lee:1993ty} through a path integral formulation but has so far never been analysed in detail except in the London limit or within the context of effective vortex-string theories. The part $\mathcal{L}^A_{\mathrm{TM}\varrho}$ describes the dynamics for a 1-form gauge field $A_{\mu}(x)$,
\begin{displaymath}
\mathcal{L}^A_{\mathrm{TM}\varrho} = \frac{-1}{4\, e^2} F_{\mu\nu}\, F^{\mu\nu} + \frac{\kappa}{6} \epsilon^{\mu\nu\rho\sigma}\, A_{\mu}\, H_{\nu\rho\sigma}\, , \nonumber
\end{displaymath}
where the conserved current $J^{\mu}(x)$ reads
\begin{equation}\label{4D_DCS-Apic-Current_1}
J^{\mu} = \frac{\kappa}{2} \epsilon^{\mu\nu\rho\sigma}\, \partial_{\nu}\, B_{\rho\sigma}\, , \quad \partial_\mu J^\mu = 0 \, .
\end{equation}

Our process of dualisation then proceeds from the definition of the first order Lagrangian density from (\ref{def:4D-TMGT+Higgs_xi=1}),
\begin{eqnarray}
\mathcal{L}^{4\mathrm{m}}_{\mathrm{TM}\varrho} & = & 
  - \frac{e^2}{4}\, E_{\mu\nu}\, E^{\mu\nu} + \frac{1}{2} \epsilon^{\mu\nu\rho\sigma}\, \partial_\mu A_\nu\, E_{\rho\sigma} \nonumber\\
& + &\frac{\eta^2}{2}\, \varrho^2\, G_\mu\, G^\mu + \frac{1}{2} \epsilon^{\mu\nu\rho\sigma}\, \partial_{\mu} B_{\nu\rho}\, G_{\sigma} \nonumber\\
& + & \frac{\kappa}{6} \epsilon^{\mu\nu\rho\sigma}  A_{\mu}\, H_{\nu\rho\sigma} + \mathcal{L}_{\varrho} \nonumber \, . 
\end{eqnarray}
In combination with the equations of motion defining the Gaussian integration, the physical variable $E_{\mu\nu} (x)$ corresponds to the electromagnetic fields,
\begin{displaymath}
 E_{\alpha\beta} = \frac{1}{e^2}\, \eta_{\alpha\mu}\, \eta_{\beta\nu}\, \epsilon^{\mu\nu\rho\sigma}\, \partial_{\rho} A_{\sigma}\, , 
\end{displaymath}
while $G_\mu(x)$ is related to the current introduced in (\ref{4D_DCS-Apic-Current_1}),
\begin{equation}\label{4D_DCS-Apic-Current_2}
J^{\mu} = \kappa\, \eta^2\, \varrho^2\, \eta^{\mu\alpha}\, G_{\alpha}.
\end{equation}
Under the same parametrisation as the one introduced in the general case,
\begin{equation}\label{def:TP-fact_4D}
   A_{\mu} = \mathcal{A}_\mu - \frac{1}{\kappa} G_\mu\, , \quad B_{\mu\nu} = \mathcal{B}_{\mu\nu} + \frac{1}{\kappa} E_{\mu\nu}\, ,
\end{equation}
the dual Lagrangian density is again factorised modulo a surface term,
\begin{displaymath}
 \mathcal{L}^{\mathrm{fac}}_{\varrho} = \mathcal{L}_{\mathrm{MHP}} + \frac{\kappa}{6} \epsilon^{\mu\nu\rho\sigma}  \mathcal{A}_{\mu}\, \mathcal{H}_{\nu\rho\sigma},
\end{displaymath}
into a topological $BF$ sector of gauge variant variables and a dynamical sector of physical variables, $\mathcal{L}_{\mathrm{MHP}}$:
\begin{eqnarray}
\mathcal{L}_{\mathrm{MHP}} &=& -\frac{e^2}{4} E_{\mu\nu}\, E^{\mu\nu} + \frac{\eta^2}{2}\, \varrho^2\, G_\mu\, G^\mu + \mathcal{L}_{\varrho} \nonumber\\ 
& + & \frac{1}{2\, \kappa} \epsilon^{\mu\nu\rho\sigma}\, \partial_\mu G_\nu\, E_{\rho\sigma} \label{4D-TMGT+Higgs_phys}\, .
\end{eqnarray}
This Lagrangian density
is nothing other than the first order formulation of the Maxwell-Higgs model expressed in terms of physical variables (see \cite{Govaerts:2000ys,Chernodub:2008rz} for a review). In this sense, $\varrho(x)$ plays the role of the Higgs field provided that the potential in (\ref{def:Free_Higgs(2)}) be chosen of the form:
\begin{displaymath}
   V\left(\varrho^2\right) = \frac{\tilde{\mu}^2}{2}\, \varrho^2 + \frac{\lambda}{4}\, \varrho^4 \, ,
\end{displaymath}
where $\tilde{\mu}^2\!<\! 0$ and $\lambda\!>\! 0$, to recover the ``Mexican hat-shaped'' quartic potential. Then the Higgs field possesses a non vanishing vacuum expectation value,
\begin{displaymath}
   \langle \varrho \rangle = v = \sqrt{\frac{-\tilde{\mu}^2}{\lambda}} \neq 0 \, .
\end{displaymath}

At this stage, the usual formulation of the MH Lagrangian is recovered through a procedure following the example of the Stueckelberg mechanism, thus requiring a non trivial extension of the gauge content. After the transformation (\ref{def:TP-fact_4D}), $G_\mu(x)$ obeys the following equation of motion associated to the physical sector of the factorised Lagrangian (\ref{4D-TMGT+Higgs_phys}),
\begin{displaymath}
\eta^2\, \varrho^2\, G_\alpha = - \eta_{\alpha\mu}\, \frac{1}{2\, \kappa}\, \epsilon^{\mu\nu\rho\sigma}\, \partial_\nu\, E_{\rho\sigma} \, .
\end{displaymath}
A conservation law naturally arises from this equation: $\partial^\mu \left(\varrho^2\, G_\mu \right)\!=\!0$ (on shell). In this sense the variable $J_\mu \propto \varrho^2\, G_\mu$ is interpreted as a current coupled with a gauge connection. This current may be that of a complex scalar field of which the Higgs field $\varrho(x)$ is the radial part. However the connection may not be the original $A_\mu(x)$ field for the reasons already mentioned above.
Thus a new connection $\tilde{A}_\mu(x)$ is introduced, which allows to redefine $G_\mu(x)$ as
\begin{equation}\label{def:4D-TMGT+Higgs_current}
   \kappa\, \eta^2\, G_\mu = - \frac{1}{\varrho^2} J_\mu = \kappa^2\, \eta^2\, \left(\tilde{A}_\mu - \partial_\mu \theta \right) \, ,
\end{equation}
where the transformation of the variable $\theta(x)$ under the new $U(1)$ gauge symmetry compensates for that of the connection $\tilde{A}_\mu(x)$ in order to preserve the gauge invariance of $J_\mu(x)$, namely  $\tilde{A}'_\mu= \tilde{A}_\mu + \partial_\mu \tilde{\alpha}$, and $\theta' = \theta + \tilde{\alpha}$.
Here the local notation for the gauge transformation is used because global variables have no influence on account of the assumptions made in the present discussion.

Given the redefinition (\ref{def:4D-TMGT+Higgs_current}) the MH model in the symmetry breaking phase is recovered from the physical sector (\ref{4D-TMGT+Higgs_phys}),
\begin{eqnarray}
\mathcal{L}^4_{\mathrm{MH}\varrho \theta} & = & -\frac{1}{4\, e^2}\, \tilde{F}_{\mu\nu}\, \tilde{F}^{\mu\nu}  \nonumber\\
& + & \frac{1}{2} \left| \partial_\mu \varrho - \mathrm{i}\, \kappa\, \eta\, \varrho\, \left(\tilde{A}_\mu - \partial_\mu \theta \right)\right|^2 - V(\varrho^2) \, . \nonumber
\end{eqnarray}
Therefore, the total gauge embedded action dual to (\ref{def:4D-TMGT+Higgs}) through TP-factorisation is decoupled into the $U(1)$ Maxwell-Higgs action and a topological $BF$ action,
\small
\begin{displaymath}
   S\left[\tilde{A},\mathcal{A},\mathcal{B},\varrho,\theta\right] = S_{\mathrm{MH}}\left[\tilde{A},\varrho,\theta\right] + S_{BF}\left[\mathcal{A},\mathcal{B}\right] + \int \mathrm{ST} \, .
\end{displaymath}
\normalsize
This action is invariant under three independent classes of finite abelian gauge transformations acting in either the $\mathcal{A}$-, the $\mathcal{B}$- or the ($\tilde{A}$,$\theta$)-sector. This latter restored transformation allows to define a complex scalar field of which the polar parametrisation follows in terms of the real scalar fields $\varrho(x)$ and $\theta(x)$:
\begin{displaymath}
   \phi(x) = \frac{1}{\sqrt{2}}\, \varrho(x)\, \mathrm{e}^{\mathrm{i} \kappa \eta\, \theta(x)} \, .
\end{displaymath}
Thus the MH Lagrangian density before symmetry breaking is recovered:
\begin{equation}\label{def:Abelian_Higgs_Lag}
   \mathcal{L}_{\mathrm{MH}} = -\frac{1}{4\, e^2}\, \tilde{F}_{\mu\nu} \tilde{F}^{\mu\nu} + \left| \tilde{\mathrm{D}}_\mu \phi \right|^2 - V\left(2\, |\phi|^2\right) \, .
\end{equation}
where the field strength tensor $\tilde{F}_{\mu\nu}(x)$ and the covariant derivative,
\begin{displaymath}
   \tilde{\mathrm{D}}_\mu \phi = \partial_\mu \phi - \mathrm{i}\, \kappa\, \eta\, \tilde{A}_\mu\, \phi \, ,
\end{displaymath}
are defined from the connection $\tilde{A}(x)$.  The dual formulation of the MH model is extensible to any dimension, notwithstanding the non renormalisable character of the scalar field quartic potential,
\begin{displaymath}
   V\left[2\, |\phi|^2\right] = \tilde{\mu}^2\, |\phi|^2 + \lambda\, |\phi|^4 \, ,
\end{displaymath}
in more than four spacetime dimensions.

\section*{Conclusions and Perspectives}

Some nodes of the network of duality relations in Fig.\ref{fig:Duality_Mass-generation} that apply for
abelian U(1) gauge theories and their mass generating mechanisms have been partially analysed in the literature.
However our approach is different. We have established that these equivalences are true modulo a topological $BF$ term and, as usual, a specific gauge embedding procedure. Although they share the same local formulation
in terms of (physical) gauge invariant fields, the gauge symmetry content of the different abelian mass generation mechanisms of Fig.\ref{fig:Duality_Mass-generation} differs dramatically. It implies that these theories share a common dynamics but are globally distinct as soon as topological effects appear. Thus we may rather admit that two gauge theories are locally dual in their physical sector but are not dual if their gauge symmetry content is considered, instead of establishing dualities at any cost. This specific feature comes to the fore in the case of TMDGT where the topological
sector of gauge variant variables factorises out from the physical sector, independently of any gauge fixing procedure. Actually this topological term was until now swamped by a mass of successive procedures of gauge embeddings and/or gauge fixings characteristic of the dualisation techniques previously introduced in the literature.  

The dielectric lagrangian (\ref{def:4D-TMGT+Higgs}) has never been studied so far in the context of condensed matter and offers a possible dual formulation of effective superconductivity without symmetry breaking. From a local 
perspective, the squared scalar field $\varrho^2(x)$ measures the local density of Cooper pairs while the $BF$ term describes a topological order in the long wavelength limit, see \cite{Hansson:2004wca}. Our duality network makes
a clear link between different effective descriptions of superconductivivity and Josephon junction arrays, see \cite{Hansson:2004wca,Diamantini:1995yb,Choudhury:2015rua}, which are recovered in the various limits of Fig.\ref{fig:Duality_Mass-generation}, also including the abelian Higgs-model with an extra topological $BF$ term\cite{Nielsen:1973cs,Hansson:2004wca}. Its generalisation also provides a fresh perspective on the higgsless superconductivity of \cite{Diamantini:2014iqa}.

However most of the other works mentioned above deal with compact gauge groups to invoke the appearance of string or monopole-like defects in 3+1 dimensions. In our network of dualities, the $BF$ term turns out to play a pivotal role as soon as topological effects are taken into account. In fact any zero of the scalar field(s) on some subset of space is associated with the existence of a topological defect localised on this subset. In particular in the dual Maxwell-Higgs model (\ref{def:4D-TMGT+Higgs}), the topological sector carries the topological content, related to vorticity, of dual Nielsen-Olesen vortices, as will be discussed elsewhere. Therefore vortex string solutions arises per se, rather than being introduced ``by hand" through the coupling of the gauge fields with the worldsheet \cite{Hansson:2004wca,Lee:1993ty} or requiring compact gauge groups as in \cite{Diamantini:1995yb,Diamantini:2014iqa}.
Furthermore, the straightforward generalisation (\ref{def:TMGT+Dielectric_fields}) to fields of any tensorial rank whatever the number of spacetime dimensions suggests a new avenue towards the construction of generalised topological defects, currently under investigation. The extension of this network of dualities to topological defect solutions may offer a new mathematical framework for the detection of topological dark matter, see \cite{Derevianko:2013oaa}. 

\section*{Acknowledgement}

This work was supported by the Belgian Federal Office for Scientific, Technical and Cultural Affairs through the Interuniversity Attraction Pole P6/11.


\end{document}